\documentclass[aps,prl,showpacs,notitlepage,floatfix,twocolumn,superscriptaddress,nofootinbib]{revtex4-1}
\usepackage{amssymb}
\usepackage{amsmath}
\usepackage{amsfonts}
\usepackage{bbm}
\usepackage{graphicx}
\usepackage{braket}
\usepackage[colorlinks,linkcolor=blue,anchorcolor=blue,citecolor=blue,urlcolor=blue]{hyperref}
\usepackage[dvipsnames]{xcolor}
\usepackage[titletoc]{appendix}
\usepackage{tikz,pgf}
\usepackage{enumerate}
\usetikzlibrary{knots}

\usepackage[normalem]{ulem}

\graphicspath{{Figures/}}

\def\bea{\begin{equation}\begin{aligned}}
\def\eea{\end{aligned}\end{equation}}
\def\tr{\text{tr}}
\def\epr{\text{EPR}}
\def\eig{\text{EIG}}

\begin{document}
\title{Long-Range Bell States from Local Measurements and Many-Body Teleportation without Time-Reversal}

\author{Lakshya Agarwal}
\thanks{These two authors contributed equally}
\affiliation{Department of Physics \& Astronomy, Texas A\&M University, College Station, Texas 77843, USA}

\author{Christopher M. Langlett}
\thanks{These two authors contributed equally}
\affiliation{Department of Physics \& Astronomy, Texas A\&M University, College Station, Texas 77843, USA}

\author{Shenglong Xu}
\affiliation{Department of Physics \& Astronomy, Texas A\&M University, College Station, Texas 77843, USA}

\begin{abstract}
In this work, we study quantum many-body teleportation, where a single qubit is teleported through a strongly-interacting quantum system, as a result of a scrambling unitary and local measurements on a few qubits.
Usual many-body teleportation protocols require a double copy of the system, and backward time evolution, we demonstrate that teleportation is possible in the 2D spin-$1/2$ XY model, without these constraints.
The necessary long-range entanglement for teleportation is generated from the model hosting special eigenstates known as rainbow scars.
We outline a specific protocol for preparing this highly entangled state by evolving a product state and performing iterative measurements on only \textit{two} qubits with feedback control.

\end{abstract}
\maketitle

Advancements in coherent control over quantum many-body systems provide an exciting opportunity to study rich non-equilibrium phenomena~\cite{georgescu2014quantum,preskill2018quantum,altman2021quantum}.
Generic initial states thermalize under unitary dynamics~\cite{deutsch1991quantum,srednicki1994chaos,polkovnikov2011colloquium}.
Despite the global time-evolved state remaining pure, locally the state becomes mixed and approaches the thermal ensemble from the rapid entanglement growth.
If one prepares an initial state on a qubit of a large system and lets the system evolve, the information contained in the qubit is not retrievable via measurements on the prepared qubit, or any local subset of qubits, due to thermalization.
While that bit of quantum information is never lost -- one can always trivially act $U^\dagger$ on the entire output state to get it back, the information spreads to non-local degrees of freedom, making retrieval difficult, known as scrambling~\cite{sekino2008fast,hayden2007black,shenker2014black}.

Remarkably, Hayden and Preskill~\cite{hayden2007black} demonstrated that the information is retrievable under the right conditions~\cite{yoshida2017efficient}.
While the original motivation of the work is to study quantum information dynamics inside a black hole, the Hayden-Preskill protocol and its generalizations~~\cite{gao2017traversable,maldacena2017diving,brown2019quantum,nezami2021quantum,schuster2021many} extend to generic quantum many-body systems~\cite{you2018entanglement,cheng2020realizing,plugge2020revival,yoshida2021decodingClifford,decoherencetele}, and have been realized experimentally~\cite{landsman2019verified,blok2021quantum,wang2021verifying}.
These many-body teleportation protocols aim at designing simple quantum operations, including unitary evolution and measurements, to teleport Alice's original qubit to Bob through a strongly interacting quantum many-body system. This differs from quantum state transfer in non-interacting spin chains~\cite{bose2003quantum,christandl2004perfect,yao2011robust}.
\begin{figure}
     \centering
     \includegraphics[width=0.9\columnwidth]{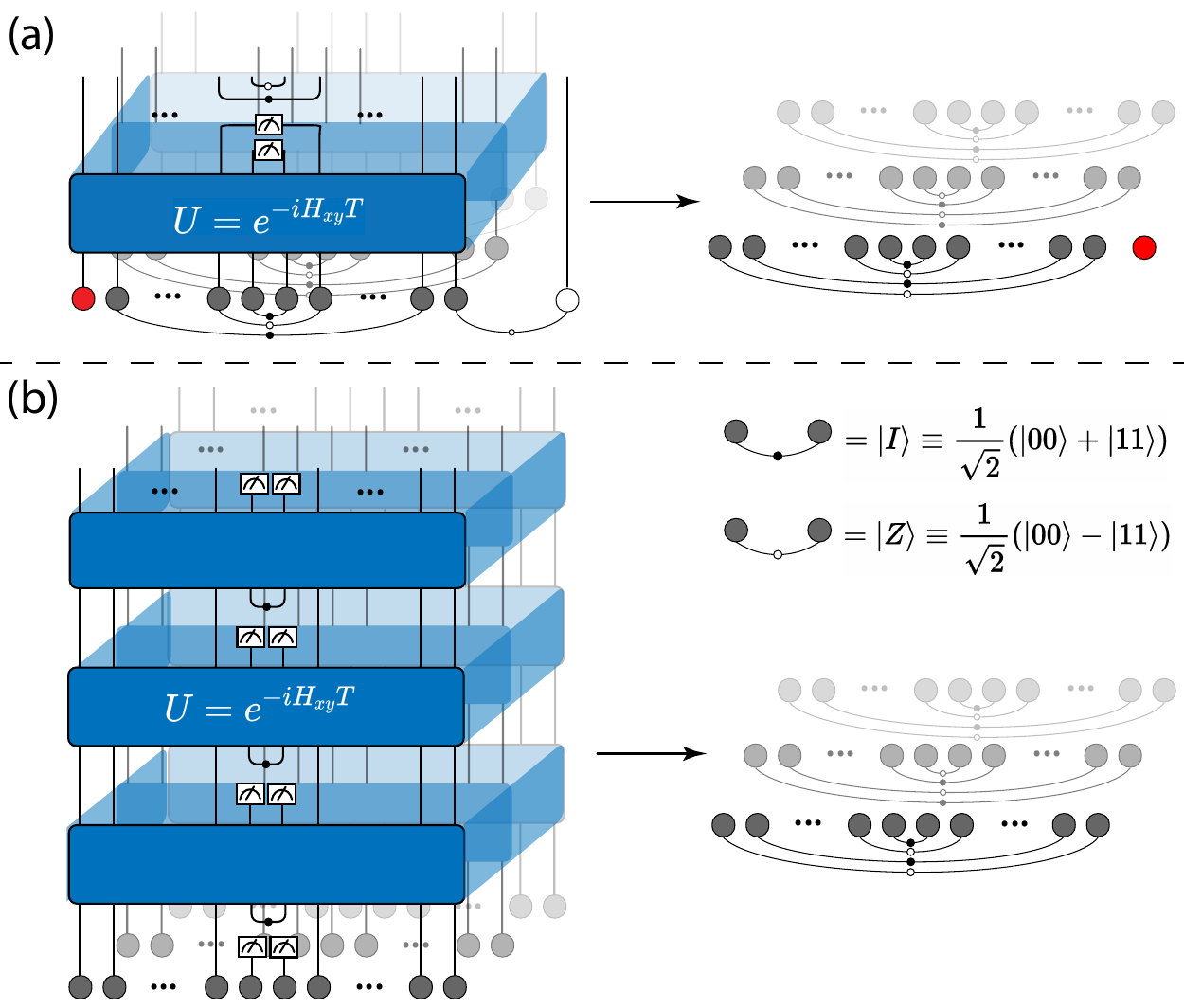}
    \caption{Summary of the main results. (a) The teleportation protocol. The initial quantum state on Alice's qubit~(red), scrambled by the unitary dynamics, is teleported to the auxiliary qubit~(white) on the other side of the system by a few Bell measurements.  (b) The state engineering protocol. The initial state required in (a) can be engineered from any initial state by scrambling unitary dynamics and periodic measurements on two central qubits with feedback control. In both protocols, only $\mathcal{O}(1)$ qubits are measured.}
     \label{fig:summary}
\end{figure}
 
Many-body teleportation generalizes conventional teleportation based on an entangled pair of qubits~(see the SM).
There are two steps, encoding and decoding the information.
In the encoding step, a single bit of quantum information spreads across the complex degrees by the unitary $U$. 
On the other hand, the decoding step concentrates the information at another qubit across the system.
Existing many-body teleportation schemes require a long-range pair-wise entangled initial state between the system and a double copy.
In addition, the usual decoder~\cite{yoshida2017efficient} contains a unitary operation $U^*$ which effectively reverses time.
The teleportation fidelity is closely related to the out-of-time ordered correlator, which is measured by directly engineering $U^*$~\cite{garttner2017measuring,mi2021information, braumuller2022probing,li2017measuring,wei2018exploring, sanchez2020perturbation} or using randomized measurements without time reversal~\cite{vermersch2019probing,nie2019detecting,joshi2020quantum}. 
However, the current scheme of direct many-body teleportation requires the long-range entangled state, a double copy of the system and time reversal, and is so far limited to small systems.

In this work, we circumvent the aforementioned obstacles.
First, we show that many-body teleportation can occur in more general quantum many-body systems without a double copy or time-reversal.
The simplest example is the homogeneous spin-$1/2$ XY model on a 2D lattice, which has been realized in various quantum simulators~\cite{browaeys2020many,monroe2021programmable,davis2020protecting,xu2020probing}. We show that the scrambling dynamics and $\mathcal{O}(1)$ Bell measurements lead to successful teleportation. These results are based on the observation that the 2D XY model, although strongly interacting, hosts special non-thermal eigenstates that are products of long-range Bell states.
These special eigenstates belong to a special class of quantum many-body scars~\cite{bernien2017probing,moudgalya2021quantum,serbyn2021quantum} referred to as rainbow scars~\cite{langlett2022rainbow, Wildeboer2022Quantum}.
Second, unitary dynamics combined with repetitive measurements of only two qubits leads to robust dynamical preparation of the rainbow state, which is otherwise hidden in the many-body spectrum and inaccessible.
In contrast to other state engineering protocols~\cite{barmettler2008quantum,alkurtass2014optimal,dutta2020long,pocklington2021stabilizing,dutta2022generating}, ours does not require control of the coupling parameters, additional dissipators, or the non-interacting properties of the 1D XY chain, and is even applicable to 2D lattices.
This work, summarized in Fig.~\ref{fig:summary}, provides a simple protocol to realize quantum many-body teleportation by exploiting unitary dynamics and local measurements. Recently, there has been an increasing interest in the role of measurements in quantum many-body systems~\cite{li2018quantum,li2019measurement,skinner2019measurement,chan2019unitary,gullans2020dynamical,jian2020measurement,zabalo2020critical,choi2020quantum,li2020conformal,jian2021measurement,tantivasadakarn2021long,lavasani2021measurement,bao2021finite,ippoliti2021entanglement,noel2021observation,koh2022experimental}, most work considers local measurements on an extensive portion of the system. We demonstrate that measuring only $\mathcal{O}(1)$ qubits, combined with scrambling unitary dynamics and scars lead to rich non-equilibrium phenomena.

First, we briefly describe the Hayden-Preskill protocol and the measurement based decoder~\cite{yoshida2017efficient}. 
Consider a system of $N$ qubits $q_1,\ldots, q_N$ maximally entangled with another $N$ qubits $q_{\tilde{1}}, \ldots, q_{\tilde{N}}$ by forming $N$ EPR pairs between $q_i$ and $q_{\bar i}$.
Bob has control of $q_{\tilde{1}}, \ldots, q_{\tilde{N}}$ plus an $\mathcal{O}(1)$ number of qubits in the system, denoted by $q_{\{E\}}$.
Alice prepares an initial state $\ket{A}$ at $q_1$ believing that the information is safe after evolution by the unitary operator $u$ because the system is initialized in a fully mixed state, and $\ket{A}$ on $q_1$  will soon be lost due to thermalization, and become unrecoverable.
However, it turns out that Alice's state is recoverable through operations on only Bob's qubits.
Bob takes a new qubit $q_b$ and entangles it with $q_{\tilde 1}$ as an EPR pair, he then engineers a $u^*$ to act on $q_{\overline 1},\ldots, q_{\overline N}$.
He then performs $E$ Bell measurements on $q_{\{E\}}$ and their partners $q_{\{\overline{E}\}}$ and then post-selects the results to be an EPR state.
Graphically, the protocol is depicted by,
\begin{equation}
\begin{tikzpicture}[rounded corners = 2pt, scale=0.8, transform shape]
\draw (0.1, 1.2) -- ( 0.1, 0);
\draw (0.4, 1.2) -- ( 0.4, 0) .. controls (1.8, -0.6) and (2.2, -0.6) .. (3.6,0) -- (3.6,1.2);
\node[circle, fill=black, minimum width=0.12cm, inner sep=0]  at (2.0,-0.45) {};
\draw (0.7, 1.2) -- ( 0.7, 0) .. controls (1.8, -0.4) and (2.2, -0.4) .. (3.3,0) -- (3.3,1.2);
\node[circle, fill=black, minimum width=0.12cm, inner sep=0]  at (2.0,-0.3) {};
\node[circle, fill=black, minimum width=0.12cm, inner sep=0]  at (2.0,-0.15) {};
\node[circle, fill=black, minimum width=0.12cm, inner sep=0]  at (2.0,-0.0) {};
\node[circle, fill=black, minimum width=0.12cm, inner sep=0]  at (2.0,1.2) {};
\node[circle, fill=black, minimum width=0.12cm, inner sep=0]  at (2.0,1) {};
\node[circle, fill=black, minimum width=0.12cm, inner sep=0]  at (4.05,0) {};
\node at (1.1,0)  {$\cdots$};
\node at (2.9,0)  {$\cdots$};
\node at (1.1,0.9)  {$\cdots$};
\node at (2.9,0.9)  {$\cdots$};
\draw (1.5, 1) -- ( 1.5, 0) .. controls (1.8, -0.2) and (2.2, -0.2) .. (2.5,0) -- (2.5,1) -- (2.5,1.2) -- (1.5, 1.2) -- (1.5, 1.0);
\draw (1.8, 1) -- ( 1.8, 0) .. controls (2.0, -0) .. (2.2,0) -- (2.2,1) -- (1.8,1);
\node[circle, fill=red, minimum width=0.2cm, inner sep=0] (1) at (0.1,0) {};
\draw (3.9, 1.2) -- ( 3.9, 0) .. controls (4.05, -0) .. (4.2,0)--(4.2,1.2);
\node[rectangle, draw=black, fill=NavyBlue, thick, rounded corners=2pt, minimum width=1.9 cm,minimum height=0.5cm, anchor=south west] (U) at (0,0.25)
{$u$};
\node[rectangle, draw=black, fill=orange, thick, rounded corners=2pt, minimum width=1.9cm,minimum height=0.5cm, anchor=south west] (U) at (2.1,0.25) {$u^*$};
\end{tikzpicture}
\end{equation}
The initial state of $q_1$, its mirror partner $q_{\overline 1}$, and the auxiliary qubit $q_b$ is $\ket{A}_1(\ket{00}+\ket{11})_{\overline 1 b}/\sqrt{2}$.
The initial state on the $2N+1$ qubits in the Bell basis of $q_1$ and $q_{\overline 1}$ is,
\bea\label{eq:init}
\ket{\psi}
= \frac{1}{2}\ket{\epr}\ket{A}_b +\sum\limits_s \frac{1}{2}\sigma_1^s\ket{\epr}\sigma_b^s\ket{A}_b.  
\eea
Where $\ket{\epr}$ is a product of $N$ EPR states between $q_i$ and $q_{\overline i}$, and $\sigma^s$ are the Pauli operators. 
The first term, where the state $\ket{A}$ appears over $q_b$, is an eigenstate of the composite unitary $u\otimes u^*$~\cite{channel1}.
However, the second term is not, and the perturbation caused by the Pauli operator $\sigma_1^s$ on the first qubit spreads across the system disrupting the pairwise entanglement.
Therefore in the late-time regime, Bob can separate the first term from the others and acquire the state $\ket{A}$ by performing Bell measurements on only a few pairs of qubits and then post-select the outcome to be an EPR state.
Only a small number of post-selections is required to select the first term of Eq.~\eqref{eq:init} so that the state $\ket{A}$ is teleported to $q_b$.
Local measurements with post-selection act effectively as a long-range projection on the time-evolved state to force $q_1$ and $q_{\overline 1}$ into an EPR state, leading to teleportation just like the single-body teleportation.

From the discussion above, the choice of a factorizable unitary $U=u \otimes u^*$ ensures that the global EPR states do not thermalize.
However, this is not the only choice.
We now arrive at the general condition for the unitary dynamics $U$ to achieve many-body teleportation:
\begin{enumerate}
    \item $U$ thermalizes a typical initial state, i.e., $U$ is not integrable.
    \item $U$ hosts a special non-thermal eigenstate exhibiting long-range and pair-wise entanglement structure, such as the EPR state.
\end{enumerate}
$U=u\otimes u^*$ meets both conditions above.
However, such a unitary is challenging to engineer because it requires a duplicate copy and time rewinding the second copy. 

\textit{Teleportation without Time Reversal.}---
We now demonstrate a simple system whose dynamics satisfy both conditions above without requiring a duplicate copy or time reversal.
In particular, the XY model, a prototypical quantum many-body model realized in various quantum simulators~\cite{browaeys2020many,monroe2021programmable,davis2020protecting,xu2020probing}.
Consider the XY model on a square lattice of size $L_x \times L_y$,
\begin{equation}
H_{xy}= \sum_{\braket{ij}} \left(J_x \sigma^x_i \sigma^x_j + J_y\sigma^y_i \sigma^y_j\right),
\end{equation}
where $\braket{ij}$ stands for nearest neighbor.
We choose $J_x\neq J_y$ to avoid charge conservation, and set $J_y=1.2 J_x$. 

Now we demonstrate that this model satisfies both conditions for teleportation.
First, unlike the 1D chain, the 2D XY model is strongly correlated even under a Jordan-Wigner transformation, thus a typical state thermalizes, satisfying the first condition. 
Second, we show that the model hosts a special long-range pair-wise entangled eigenstate whereby a qubit $q_{i,j}$ forms a Bell state with its mirror partner $q_{\overline i,j}$, where $\overline i=L_x-i$.
Two of the Bell states, $\frac{1}{\sqrt 2}(\ket{00}\pm \ket{11})$ are denoted as $\ket{I}$ and $\ket{Z}$ respectively.
The eigenstate is written as,
\begin{equation}    
\label{eq:eigen}
\begin{aligned}
\ket{\eig}=\frac{1}{2^{L_x L_y/4}}\prod\limits_{i,j}\left(1 +(-1)^{i+j} S^+_{i,j}S^+_{L_x-i,j}\right )\ket{0}.
\end{aligned}
\end{equation}
Each term in the product creates a Bell pair, either $I$ or $Z$, based on the sign $(-1)^{i+j}$, forming a checkerboard pattern. 
In the SM we verify the state is an eigenstate of the XY model and identify three other closely related long-range entangled eigenstates.
The special eigenstate identified here exists for arbitrary $L_y$ and even $L_x$, built from long-range Bell states within each row.
The state identified here is different from the previously known exact eigenstates in the XY ladder~($L_y=2$) built from short-range Bell pairs on the rungs~\cite{vznidarivc2013coexistence, iadecolaladder}.
Since the special eigenstate $\ket{\eig}$ is related to $\ket{\epr}$ by single-qubit rotations, the XY model meets both conditions for many-body teleportation.

Now we present the many-body teleportation protocol in the XY model.
Begin with the eigenstate in Eq.~\eqref{eq:eigen}. 
We choose the first row, break the long-range entanglement between $q_{1,1}$ and $q_{\overline{1},1}$, and inject the state $\ket{A}$ into $q_{1,1}$ to be teleported.
Second, we take a new auxiliary qubit $q_b$ and entangle it with $q_{\overline{1},1}$.
The state, similar to Eq.~\eqref{eq:init}, in the Bell basis of $q_{1,1}$ and $q_{\bar 1,1}$, reads
\begin{equation}\label{eq:xy_init}
\ket{\psi}
= \frac{1}{2}\ket{\eig}\ket{A}_b +\sum\limits_{s=x,y,z} \frac{1}{2}\sigma_{(1,1)}^s\ket{\eig}\sigma_b^s\ket{A}_b. 
\end{equation}
Then the state is evolved for a time $t$ under the Hamiltonian which does not affect the auxiliary qubit $q_b$.
Afterward, local Bell measurements are performed on a few qubits and their mirror partners, and results are post-selected to match the pattern in $\ket{\eig}$ ($\ket{I}$ or $\ket{Z}$) to stabilize the eigenstate.
Because of thermalizing dynamics, the pairwise entanglement structure in the second term is disrupted by the perturbation $\sigma^s_{(1,1)}$ over time, thereby only the first term survives post-selection given a few measurements.
The initial state on $q_1$ is teleported to the auxiliary qubit $q_b$, from the combined effect of thermalization, the special eigenstate, and Bell measurements. The protocol can be characterized by the teleportation fidelity $F_E = \bra{A}\rho_b\ket{A}$, where $\rho_b$ is the density matrix of the auxiliary qubit given successful postselections of $E$ Bell measurements. 
Assuming that the second term thermalizes to the local fully mixed state, the probability $P_E$, of post-selecting $E$ Bell measurements, and the teleportation fidelity are
\begin{equation}\label{eq:PF}
P_E = \frac{1}{4} +\frac{1}{4}\frac{3}{4^E},\ \ F_E = 1 - \frac{2}{4^E+3}.
\end{equation}
A detailed derivation is given in SM. As the number of Bell measurements increases, the probability of post-selection decreases to 1/4 and $F_E$ saturates to 1 exponentially fast.
\begin{figure}
    \centering
    \includegraphics[width=0.45\columnwidth]{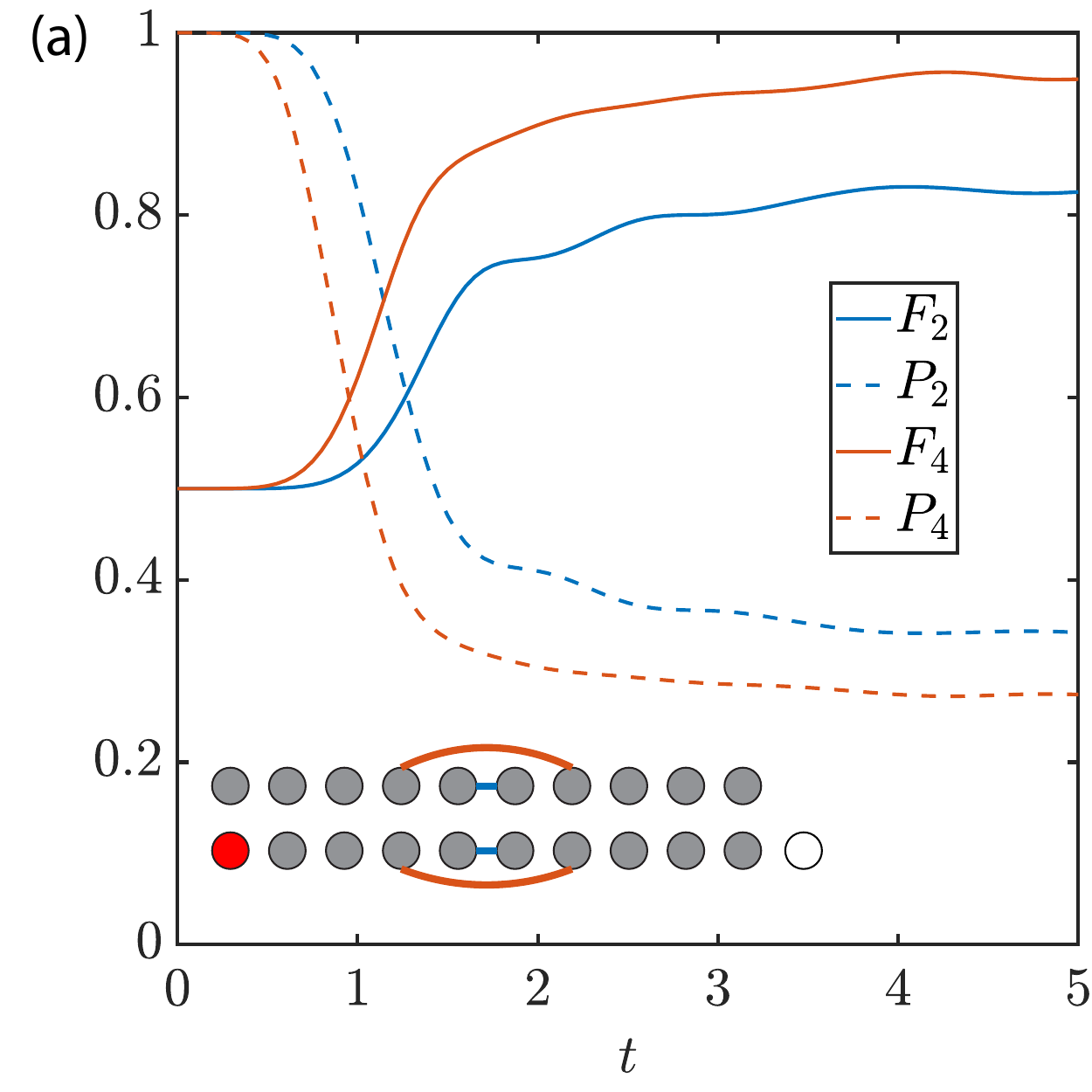}
    \includegraphics[width=0.45\columnwidth]{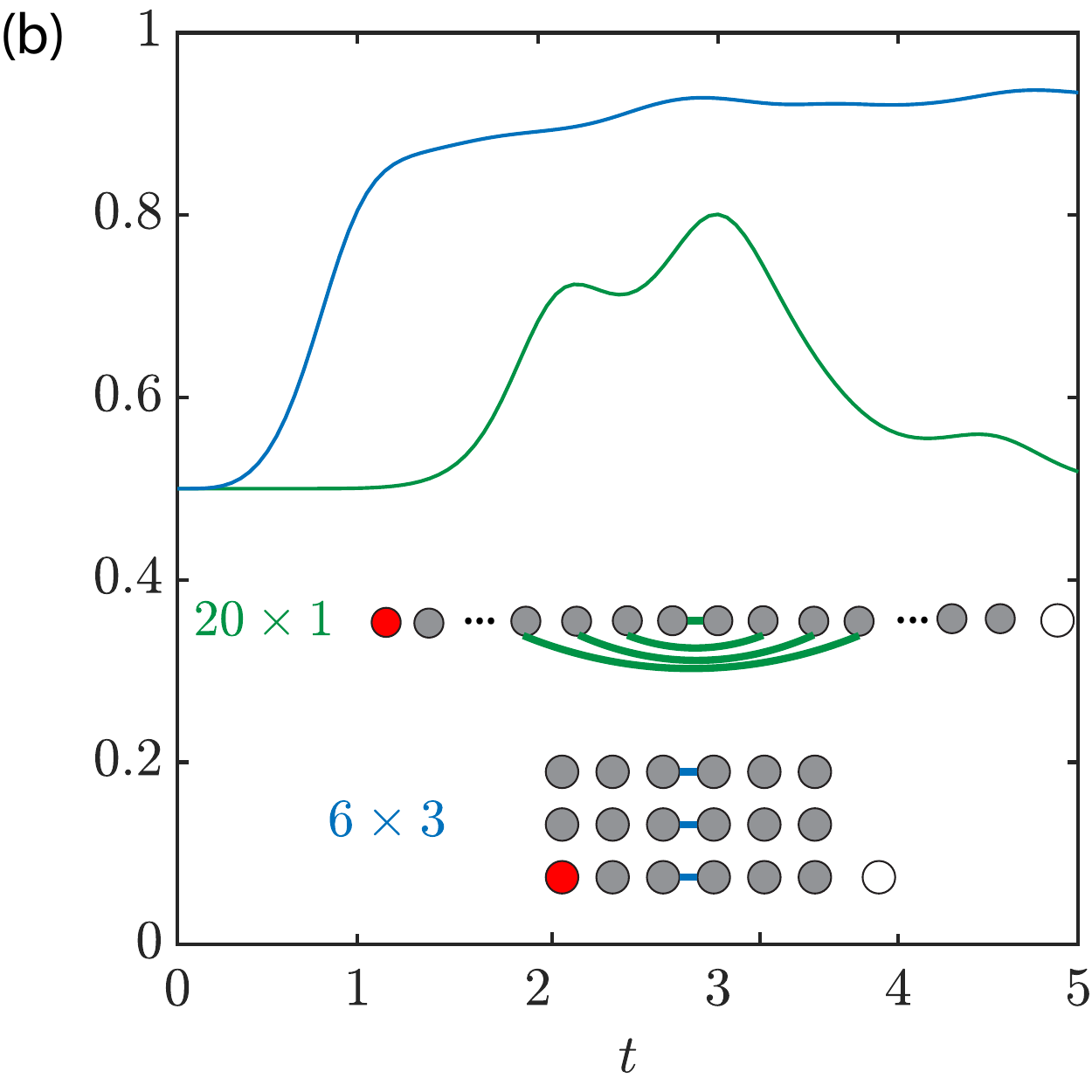}
    \caption{(a) The teleportation protocol of the XY model with a geometry of $10\times 2$ qubits. The solid and dashed lines represent the fidelity and post-selection probability, respectively. 
    In the first case, two center qubit pairs, shown in the inset, are measured and post-selected~(blue). In the second case, two additional pairs are measured~(orange). 
    (b) We repeat the simulation for the geometries of $20\times 1$ and $6\times3$. The teleportation fails in the 1D case because it is non-interacting.}
    \label{fig:teleportation}
\end{figure}

Numerically, we simulate the teleportation protocol on a ladder~($L_y=2$)~(see SM for numerical implementation).
We emphasize that the teleportation occurs from one side of a leg to the other side whereby the ladder geometry makes the model interacting.
Fig.~\ref{fig:teleportation}(a) plots both the probability $P_E$ of successful post-selection and the teleportation fidelity $F_E$.
The inset illustrates two scenarios where either two or four pairs of qubits are measured with post-selection.
The post-selection probability decreases as time proceeds but saturates to a finite value, becoming $\sim 1/4$ when four pairs of qubits are measured, indicating that only the first term in Eq.~\eqref{eq:xy_init} is selected.
The fidelity initially stays at $1/2$ until the perturbation of $\sigma^s_{(1,1)}$ reaches the measured qubits and quickly saturates to the maximal value $F_{E}\sim 1$ for a random initial state $\ket{A}$, confirming Hayden-Preskill like teleportation in the XY model \textit{without} explicit time-reversal.
The Bell measurements are essential for the protocol, without which $\rho_b$ would remain mixed and the fidelity would stay at $1/2$.

To emphasize the necessary condition of thermalization in the teleportation procedure, we repeat the protocol for the free 1D XY chain in Fig~\ref{fig:teleportation}(b), showing that the fidelity is significantly below 1.
This is because the post-selection of the Bell measurement cannot completely rule out the second term in Eq.~\eqref{eq:xy_init} which does not thermalize in the free case. We repeat the protocol for $6\times3$ qubits, showing geometry independence.

\textit{Initial state engineering.}---
The teleportation protocol above begins with a special long-range entangled initial state $\ket{\eig}$, which is often challenging to engineer experimentally.
The second main result of this work is a state-preparation protocol that engineers $\ket{\eig}$ from an arbitrary state. 
The state-preparation protocol iteratively applies a quantum channel $\mathcal{M}$ that combines unitary time evolution of the XY Hamiltonian, measurement, and feedback control. The measurement is only on two neighboring qubits $q_c$ and $q_{\bar c}$ in the middle of the system, that form an EPR pair in $\eig$.
The protocol drives an arbitrary initial state to the steady state of $\mathcal{M}$, which is engineered to be the pure state $\ket{\eig}$, i.e., 
$\lim \limits_{n\rightarrow \infty} \mathcal{M}^n(\cdot) = \ket{\eig}\bra{\eig}$. 
This is accomplished with the following two steps in $\mathcal{M}$:
\begin{enumerate}[1:]
\label{lst:protocol}
    \item (Evolution under $H_{\text{xy}}$): Evolve the global state for a time $T$ under $H_{\text{xy}}$.
    \item (Measurement and feedback): Measure the two-qubit state of $q_c$ and $q_{\bar c}$ in the computational basis. 
    If the outcome is $\ket{11}$ or $\ket{00}$, rotate the two-qubit state into the EPR state. 
    If the outcome is $\ket{01}$ or $\ket{10}$, reset the global state to $\ket{00\cdots00}$ and then rotate the two-qubit state of $q_c$ and $q_{\bar c}$ into the EPR state.
\end{enumerate}
\noindent The first step does not affect $\ket{\eig}$ because it is an eigenstate of $H_{xy}$. The second step does not change $\ket{\eig}$ because $q_c$ and $q_{\bar c}$ in $\ket{\eig}$ already form an EPR state, and the measurement always leads to $00$ or $11$, which is rotated back into the EPR pair.
The feedback is essential to ensure that $\ket{\eig}$ is the unique steady state, without which a typical initial state would converge to the fully mixed state~(see SM for an example).
After each iteration, $q_c$ and $q_{\bar c}$ are always in the EPR state because of the feedback, and the system is described by the density matrix $\rho$ of the remaining $N-2$ qubits. In each iteration, the density matrix $\rho$ is updated by $\mathcal{M}$ as
\begin{equation}    
\begin{aligned}
\label{eq:channel}
\rho^{(n+1)} &= \mathcal{M} [\rho^{(n)}] = K_{00} \rho^{(n)} K^\dagger_{00} + K_{11} \rho^{(n)} K^\dagger_{11} \\
&+ \left(\tr \left(K_{01} \rho^{(n)} K_{01}^\dagger \right) + \tr \left(K_{10} \rho^{(n)} K_{10}^\dagger \right) \right) \rho_0
\end{aligned}
\end{equation}
Here $K_s(T)=\bra{s}e^{-iH_{xy}T}\ket{I_{c\overline c}}$ has dimensions $2^{N-2}$ and $n$ represents the number of iterations through the above steps.
The last two terms account for state reset, and $\rho_0=\ket{0\cdots0}\bra{0\cdots0}$ is on all qubits except $q_{c}, q_{\overline c}$.

\begin{figure}
    \centering
    \includegraphics[width=0.45\columnwidth]{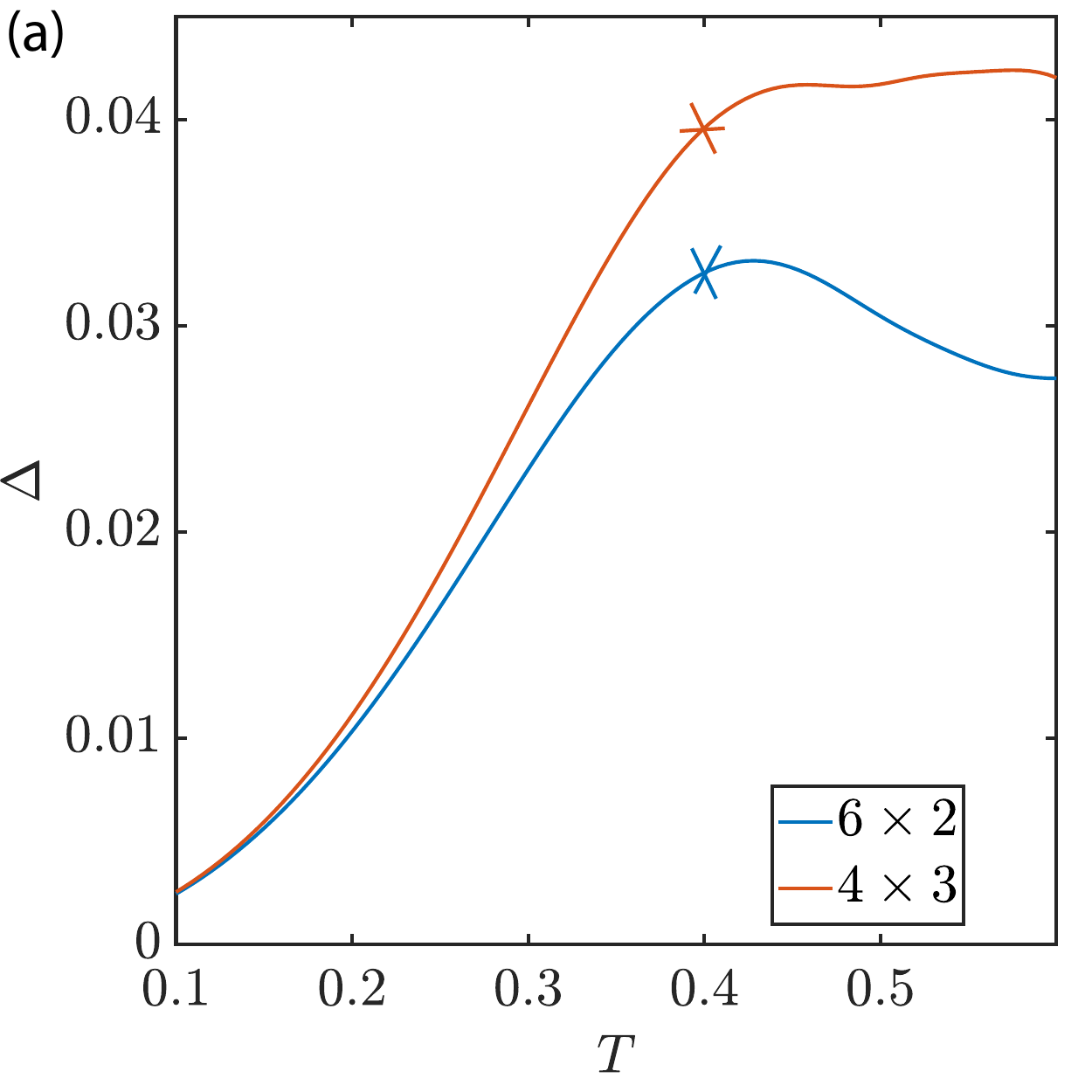}
    \includegraphics[width=0.45\columnwidth]{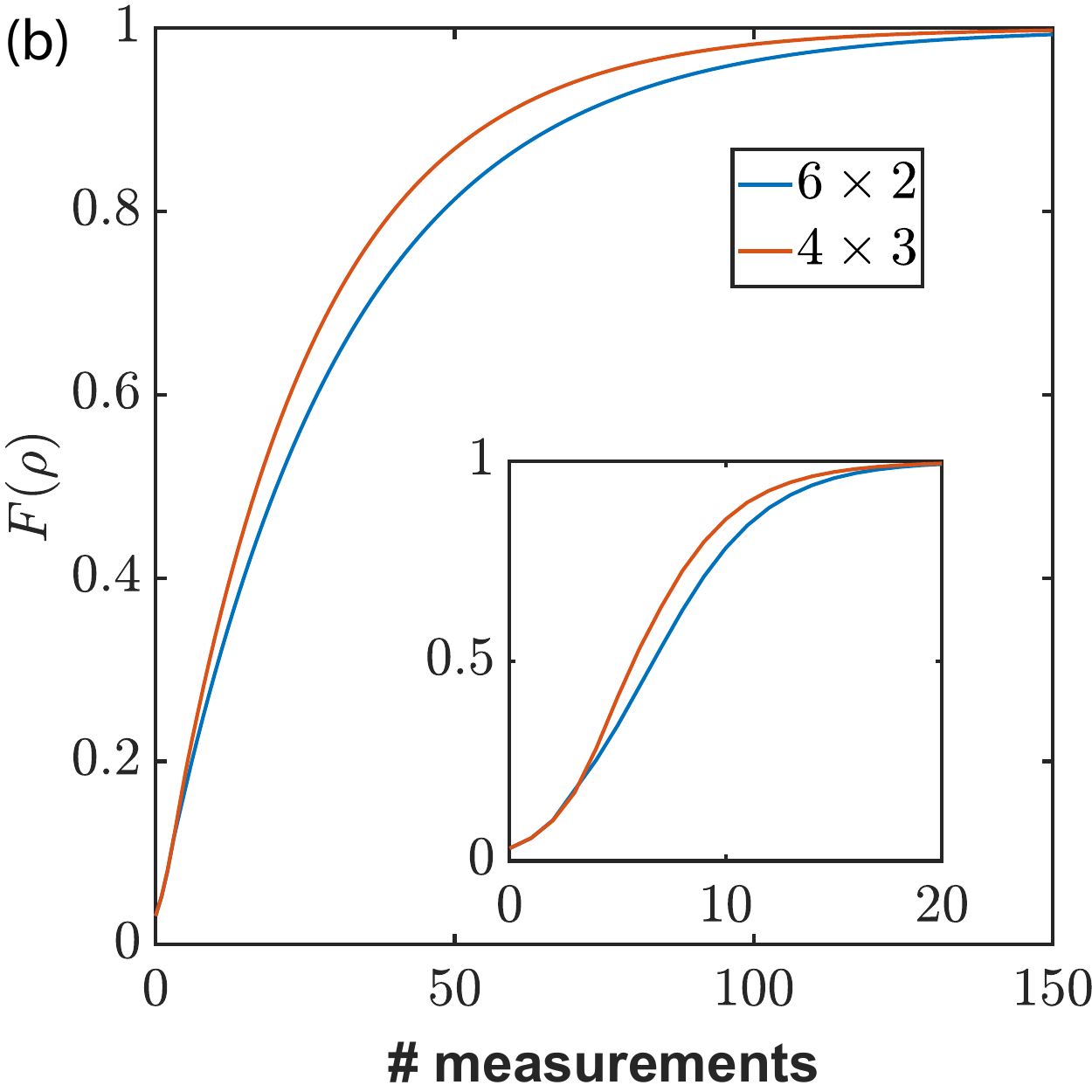}
    \includegraphics[width=0.45\columnwidth]{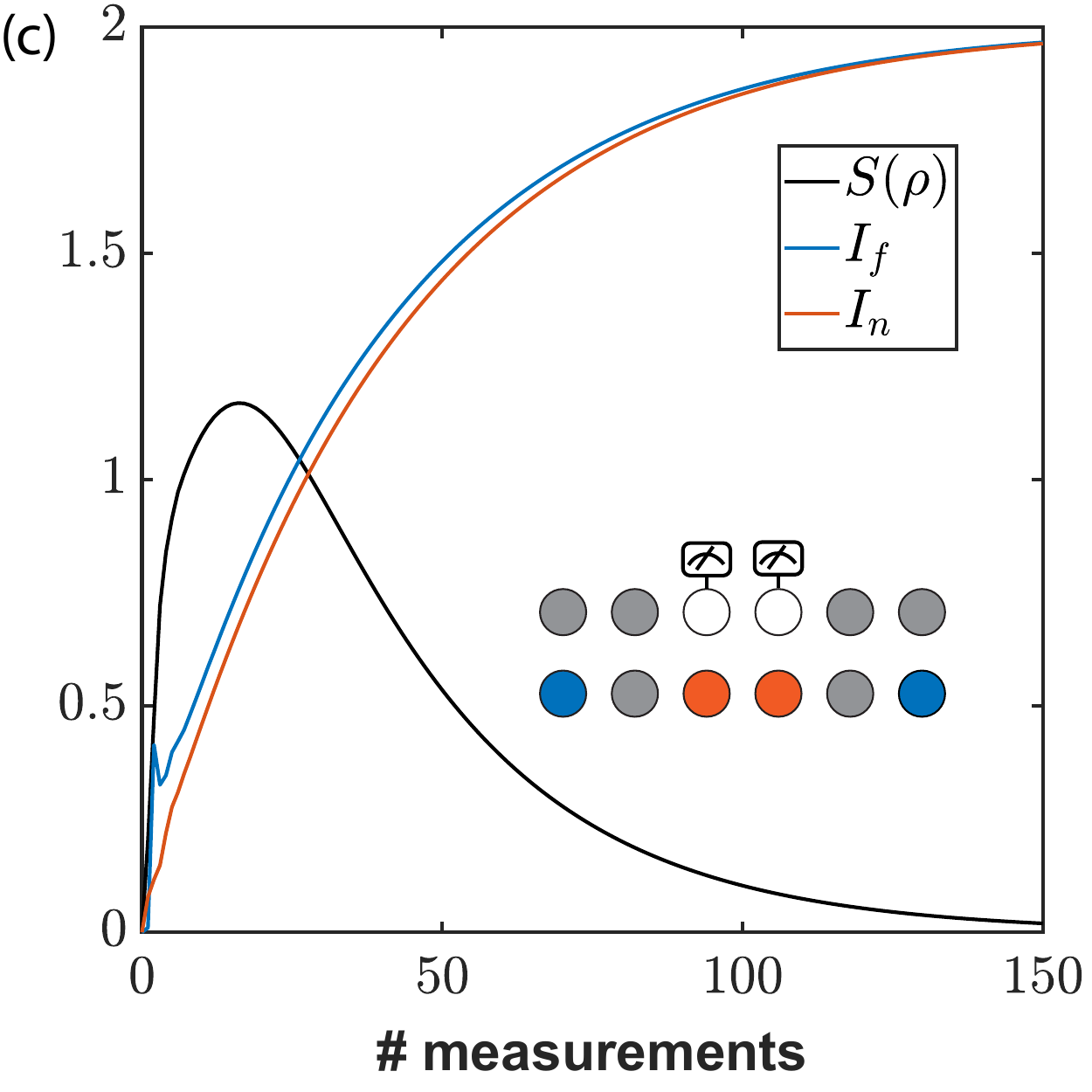}
    \includegraphics[width=0.45\columnwidth]{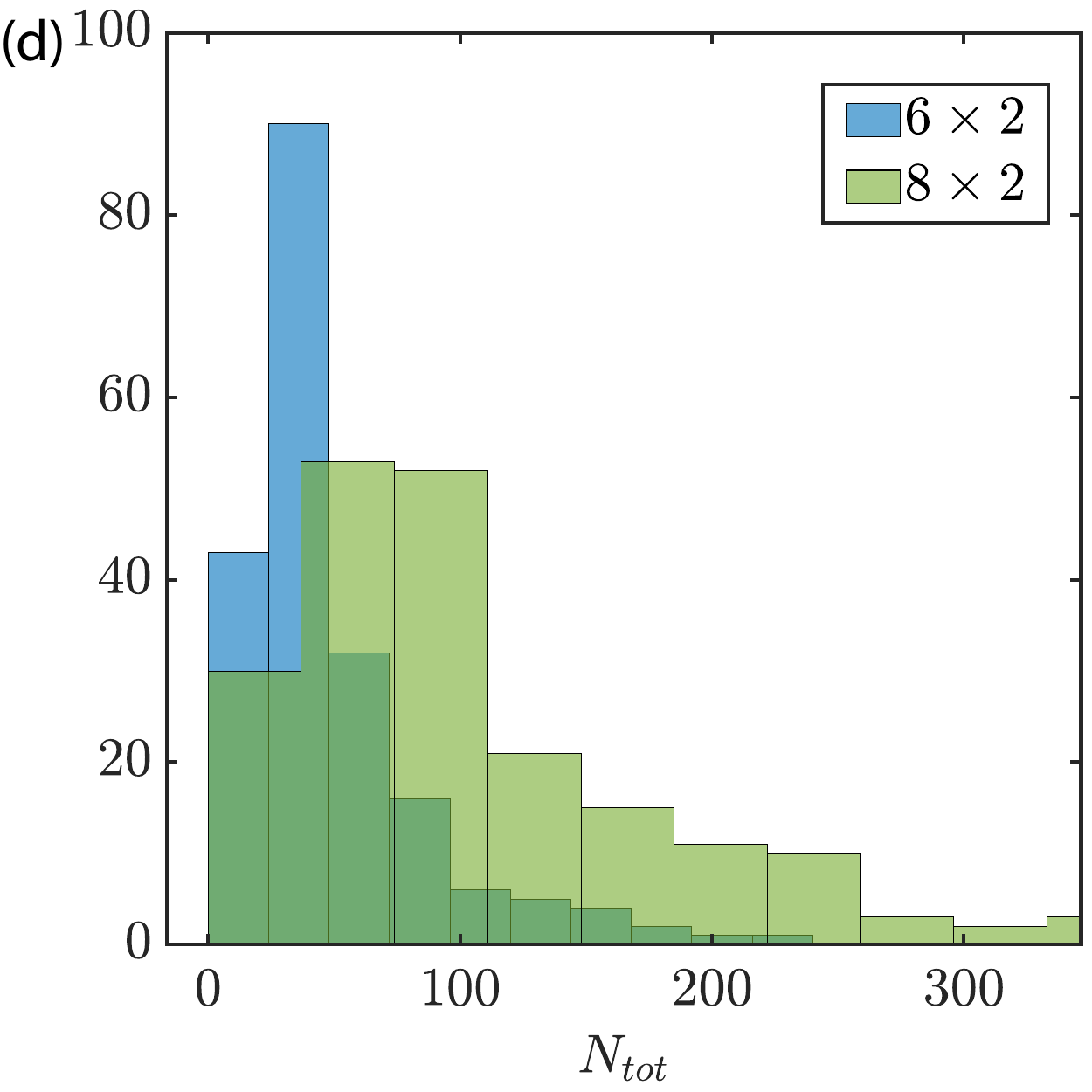}

    \caption{(a) The gap of the quantum channel in Eq.~\eqref{eq:channel}, as a function of measurement period, $T$. 
    (b) The fidelity between the density matrix $\rho$ and the target state as a function of the number of measurements. Inset: Only considering the dynamics after the final reset, $\rho$ reaches the unique steady state after $\sim 20$ measurements.
    (c) The R\'enyi entropy of the density matrix~(black), and the mutual information between the two qubits in the furthest pair~(blue) and the nearest pair~(orange) as a function of the number of measurements. (d) Histogram of the total number of measurements to reach fidelity 0.99, averaged over 200 quantum trajectories.}
    \label{fig:state_preparation}
\end{figure}

The quantum channel $\mathcal{M}$ depends on the measurement period $T$. We first characterize the spectrum gap of $\mathcal{M}$ as a function of $T$. The spectrum gap is defined as $\Delta=1-|\lambda|$ where $\lambda$ is the second largest eigenvalue of $\mathcal{M}$ and determines how fast an initial state converges to the steady state during the iteration. Fig.~\ref{fig:state_preparation}(a) plots $\Delta$ as a function of $T$ for 12 qubits in two different geometries. The finite gap confirms a unique steady state $\ket{\eig}$, and the maximal value of $\Delta$ for both geometries occurs at $T\sim 0.4$, which is chosen for the following simulations. 
Fig.~\ref{fig:state_preparation}(b) shows the fidelity $F(\rho)=\bra{\eig} \rho \ket{\eig}$ as a function of the number of measurements by directly simulating Eq.~\eqref{eq:channel}~(See SM for details). 
The fidelity exceeds 0.99 after $n=150$ iterations of $\mathcal{M}$ including reset, confirming that $\rho$ reaches the targeted pure state.
The reset procedure drastically reduces the required coherence time for the protocol, because the system only needs to maintain coherence between resets instead of all iterations. After the final reset, the dynamics are given by post-selecting the measurement outcome to be $00$ or $11$, corresponding to the first two terms in Eq.~\eqref{eq:channel} with proper normalization. The fidelity after the final reset~(Inset in Fig.~\ref{fig:state_preparation}(b)) exceeds 0.99 after 20 iterations. Given the evolution time for each iteration $T=0.4/J$, the required coherence time is as short as $8/J$.
In SM, we show that the protocol without reset also converges to the correct target state, but is far less efficient based on the gap of the quantum channel, and requires a much longer coherence time. We also show that without the feedback control, the initial state does not converge to the target state but instead thermalizes to the fully mixed state.

Measurements in our protocol cause an initial state that is evolving into a mixed state during early iterations to purify into the pure steady state at later iterations. This process is expressed in the non-monotonic behavior of the second R\'enyi entropy $S^{(2)}(\rho)=-\log_2 \tr(\rho^2)$ shown in Fig.~\ref{fig:state_preparation}(c), demonstrating the complex role of measurement in many-body dynamics.
The mutual information between a qubit $q_i$ and its mirror partner $q_{\overline{i}}$ monotonically increases to the maximal value 2, indicating Bell state formation.

Each run of the protocol depends on the outcomes of all measurements, which is called a quantum trajectory. 
The fidelity $F(\rho)$ reaching 0.99 indicates convergence to the steady state $\ket{\eig}$ for most quantum trajectories.
A more relevant quantity is the typical number of required measurements.
To study the distribution of the required number of measurements, we simulate the state-preparation protocol by stochastically sampling over all measurement outcomes using Monte Carlo~(see SM for details about numerical implementation). Simulating quantum trajectories also  allows us to probe larger system sizes because only a pure state instead of the density matrix is evolved.
Fig.~\ref{fig:state_preparation}(d) plots the histogram of $ N_{\text{tot}}$ obtained from 200 trajectories. For 12 qubits, the typical number of required measurements is 50, which is significantly less than 150 in Fig.~\ref{fig:state_preparation}(b). For 16 qubits, it typically requires 80 measurements to reach the steady state. In this case, we also checked the number of measurements required after the final reset, which is only 25.
Once the long-range entanglement is established, the state may be utilized in many-body teleportation protocol discussed in the beginning.

\textit{Discussion and outlook.}---
Many-body teleportation protocols require preparing the system in a highly entangled state along with time reversal.
In this work, we develop a state preparation protocol whereby the entangling resource is engineered through the interplay between unitary dynamics and iterative measurement on a \textit{single} pair of qubits with feedback control beginning from a simple product state.
Using this newly discovered long-range entangled eigenstate of the XY Hamiltonian, we demonstrate that many-body teleportation is achievable with \textit{only} forward time-evolution.
Through these two protocols, we have obviated the difficulties of quantum many-body teleportation, making the phenomena more amenable to current experimental platforms. Since the XY model is the underlying Hamiltonian of various quantum simulators, our protocols pave the way for generating long-range entanglement and many-body teleportation in analog and analog-digital simulators~\cite{gonzalez2021digital}. Combined with our previous works~\cite{langlett2022rainbow, Wildeboer2022Quantum} about a general recipe to construct strongly correlated Hamiltonians hosting long-range entangled eigenstates, this work greatly expands systems where quantum many-body teleportation is realizable.
When the Hamiltonian possesses further symmetries such as, $U(1)$ for $J_x=J_y$ or translation invariance, the special eigenstate is projected into each symmetry sector, leading to richer non-equilibrium phenomena, which is left for future work.
Another future direction will be teleportation using the finite-temperature thermofield double state~\cite{cottrell2019build, maldacena2018eternal,variationalTFD, brown2019quantum}. Furthermore, generalizing the state-engineering protocol of this work gives a road map to engineer other quantum many-body states with interesting entanglement structures, e.g.,\cite{sanjayent, turner2018weak,volker, schecter, chiral}.


\textit{Acknowledgement.}---We thank Alexey V. Gorshkov, Thomas Iadecola, Xiao Chen, Brian Swingle, Yi-Zhuang You, and Tianci Zhou for useful discussions. The numerical simulations in this work were conducted with the advanced computing resources provided by Texas A\&M High Performance Research Computing.

\bibliography{references}

\end{document}


\title{Supplementary Materials for "Long-Range Bell States from Local Measurements and Many-Body Teleportation without Time-Reversal"}

\author{Lakshya Agarwal}
\thanks{These two authors contributed equally}
\affiliation{Department of Physics \& Astronomy, Texas A\&M University, College Station, Texas 77843, USA}

\author{Christopher M. Langlett}
\thanks{These two authors contributed equally}
\affiliation{Department of Physics \& Astronomy, Texas A\&M University, College Station, Texas 77843, USA}

\author{Shenglong Xu}
\affiliation{Department of Physics \& Astronomy, Texas A\&M University, College Station, Texas 77843, USA}
\maketitle
\section{Conventional Quantum Teleportation}
Conventional single-body teleportation begins with a qubit that encodes the information, along with an EPR pair,
\bea
\ket{\psi_i}=
\begin{tikzpicture}[rounded corners =4pt, baseline=0.1cm]
\draw  (0,0) -- (0, 0.5);
\draw (0.5, 0.5) -- (0.5,0) -- (1, 0) -- (1, 0.5) ;
\fill [black] (0.75,0) circle (2pt);
\fill [red] (0,0) circle (3pt);

\end{tikzpicture}
\eea

A Bell measurement is performed on the first two qubits, where the outcome is post-selected to be an EPR state.
Graphically, the un-normalized state is,
\bea\label{eq:single_body}
\ket{\psi_f}=
\begin{tikzpicture}[rounded corners =4pt, baseline=0.1cm]
\draw (0,0) -- (0, 0.5)--(0.5, 0.5) -- (0.5,0) -- (1, 0) -- (1, 1) ;
\draw (0,1) -- (0, 0.75) -- (0.5, 0.75) -- (0.5, 1); 
\fill [black] (0.75,0) circle (2pt);
\fill [black] (0.25,0.5) circle (2pt);
\fill [black] (0.25,0.75) circle (2pt);
\fill [red] (0,0) circle (3pt);
\end{tikzpicture}
=\frac{1}{2}\ \ 
\begin{tikzpicture}[rounded corners =4pt, baseline=0.1cm]
\draw (0,0) -- (0, 0.5)--(0.5, 0.5) -- (0.5,0) -- (1, 0) -- (1, 1) ;
\draw (0,1) -- (0, 0.75) -- (0.5, 0.75) -- (0.5, 1); 
\fill [black] (0.25,0.75) circle (2pt);
\fill [red] (0,0) circle (3pt);
\end{tikzpicture}
\eea
It is clear that after the protocol, the initial state of the first qubit appears on the last qubit.
The factor of $1/2$ above is the normalization, indicating that the probability of the successful post-selection is $1/4$.
The above simple protocol is the stochastic version of single-body teleportation, which can be generalized to the many-body case. Of course, in the conventional protocol, the post-selection can be avoided. There are 4 possible outcomes of the Bell measurements, which tells how the state of the third qubit is related to the initial state of the first qubit. Therefore, based on the measurement outcome, one can apply the corresponding Pauli operator to the third qubit to retrieve the state. In the original story, Alice owns the first two qubits and Bob owns the third qubit which can be far away. Alice then performs the Bell measurement on the first two qubits and tells Bob the result over the telephone so that Bob can get Alice's state by applying the Pauli operator to his qubit.

There also exist many-body teleportation protocols that only include unitary operations and thus is free of post-selection, which is based on Grover's search and in the simplest case, can be understood as the SWAP gate.

\section{Scrambling in the Hayden-Preskill Protocol.}
This section provides additional computations in the conventional Hayden-Preskill formalism, and connects it to the teleportation protocol in the XY model presented in the main text. 

We will first begin with a basic identity regarding operator growth due to unitary dynamics.
Given an operator $O_i$ at time $t=0$, it can evolve into a more complicated operator at time $t$, under generic unitary dynamics:
\bea
O_i(t) = \sum_{A} c_{A} W_{A}
\eea
where $c_A$ are the coefficients of the operators $W_A$, which represent the operator-basis for the chosen model. 
For qubit systems, a convenient basis choice are Pauli strings. 
Under unitary evolution, all simple initial operators except the identity operator evolve into more complicated operators, where at late times all Pauli strings have roughly equal probability of appearing. This is the essence of scrambling dynamics. 
Given that the dynamics are unitary, one can easily check that the following holds:
\bea
\label{Eq:wavefunction}
\text{Tr}(O^{\dagger}(t) O(t)) = \text{const.} \implies \sum_A |c_A|^2 = \text{const.}
\eea

Operator dynamics of $n$ qubits can be interpreted as state dynamics of $2n$ qubits through operator-state duality, which can be roughly viewed as `flipping' the bra to a ket~\cite{channel1}:
\bea
O = \sum_{k,l} O_{k,l} \ket{k}\bra{l} \implies \ket{O} = \sum_{k,l} O_{k,l} \ket{k} \otimes \ket{l}
\eea
Under Heisenberg evolution, we have
\begin{equation}
    O(t) = U O U^\dagger =\sum\limits_{k,k',l,l'} U_{k,k'} U^*_{l,l'}O_{k',l'} \ket{k}\bra{l}
\end{equation}
We use $UOU^\dagger$ instead of $U^\dagger O U$ to be consistent with the Hayden Preskill protocol. Using the operator to state mapping, we see that the unitary dynamics on the state side is controlled by a decoupled unitary $U\otimes U^*$, 
\begin{equation}
    \ket{O(t)} = U\otimes U^* \ket{O}
\end{equation}
The operator-state mapping is more manifest through the following graph,
\bea
\begin{qtex}[0.5]
\grid{4}{1;4,-2}
\gate[fill=NavyBlue][center:$U$][4][1.5]{1,1}
\gate[fill=Orange][center:$U^\dagger$][4][1.5]{1,-1.5}
\gate[fill=red][][1]{1,0}
\end{qtex} \longrightarrow
\begin{qtex}[0.5]
\grid{2}{1;8,0}
\connect[rounded corners=0.2][-0.2]{1,0}{8,0}
\connect[rounded corners=0.2][-0.2]{2,0}{7,0}
\connect[rounded corners=0.2][-0.2]{3,0}{6,0}
\connect[rounded corners=0.2][-0.2]{4,0}{5,0}
\gate[fill=NavyBlue][center:$U$][4][1.5]{1,1}
\gate[fill=Orange][center:$U^*$][4][1.5]{5,1}
\gate[fill=red][][1]{4.5,-1.4}
\end{qtex}
\eea

Furthermore, the identity operator is mapped to the state of $n$ EPR pairs, dubbed as $\ket{\epr}$
\bea
&\frac{I}{2^{n/2}} = \frac{1}{2^{n/2}} \sum_{i} \ket{i}\bra{i} \implies \ket{\epr} = \frac{1}{2^{n/2}} \sum_{i} \ket{i} \otimes \ket{i}
\eea
Since the identity operator remains static under Heisenberg evolution, the EPR state is an eigenstate of the unitary $U\otimes U^*$ on the state side
\bea
&U \bigg(\frac{I}{2^{n/2}}\bigg) U^{\dagger} = \frac{I}{2^{n/2}} \implies  (U \otimes U^*) \ket{\epr} = \ket{\epr}.
\eea

Graphically, the initial state of the Hayden-Preskill protocol as described in the main text is given by
\bea
\ket{\psi} =
\begin{qtex}[0.5]
    \grid{1}{0;9,0}
    \qubit[draw=red][below:$A$]{0,0}
    \connect[rounded corners=0.2][-0.2]{1,0}{6,0}
    \connect[rounded corners=0.2][-0.2]{2,0}{5,0}
    \connect[rounded corners=0.2][-0.2]{3,0}{4,0}
    \connect[rounded corners=0.2][-0.2]{7,0}{8,0}
    \node at (4,0.7) {$b$};
    \node at (0,0.7) {$1$};
    \node at (3.5,0.7) {$\bar 1$};
    \node at (1.75,0.4) {$\cdots$};
    \ctrl{3.5,-1;3;0.4}
    \ctrl{7.5,-0.2}
\end{qtex}
\eea
where the first qubit is prepared in a state $A$ to be teleported, and Bob's qubit $b$ forms an EPR with the mirror partner of the first qubit.
Expanding the state in the Bell basis of the first qubit and its mirror partner, we get
\bea\label{eq:init}
\ket{\psi}= \frac{1}{2}\ket{\epr}\ket{A}_b +\sum\limits_{s=x,y,z} \frac{1}{2}\sigma_1^s\ket{\epr}\sigma_b^s\ket{A}_b
\eea
The two terms above are orthogonal. 
The time-evolved state becomes,
\bea
&\ket{\psi(t)}=\frac{1}{2}\ket{\psi_0} + \frac{\sqrt 3}{2}\ket{\psi_1(t)}\\
&\ket{\psi_0}=\ket{\epr}\ket{A}_b, \quad \ket{\psi_1(t)} = \sum\limits_s \frac{1}{\sqrt{3}}\sigma_1^s(t)\ket{\epr}\sigma_b^s\ket{A}_b.
\eea
The states $\ket{\psi_0}$ and $\ket{\psi_1}$ remain orthogonal at all times because of unitary time-evolution.
The state $\ket{\psi_0}$ remains unaffected as it is an eigenstate. After the scrambling time, the state $\ket{\psi_1(t)}$ is scrambled, and the corresponding density matrix $\ket{\psi_1(t)}$ of a local region $E$ will approach the fully mixed state,
\bea
\rho^E_1(t) = \text{tr}_{\bar E} \ket{\psi_1(t)}\bra{\psi_1(t)} \rightarrow \frac{1}{\text{dim} E} I^E,
\eea
assuming no conservation laws in the system.

Bob can then only select $\ket{\psi_0}$ by Bell measurements on $E$ pairs of qubits and post-select the results to be $E$ EPR states. The motivation of the Bell measurement is only to select the first term from the full state. Denote the projector on $E$ EPR state as $p_E$, and we have $p_E \ket{\psi_0} = \ket{\psi_0}$. Then the post-selection probability is 
\bea
P_E = \langle \psi(t)|p_E | \psi(t) \rangle = \frac{1}{4} + \frac{3}{4} \bra{\psi_1(t)} p_E \ket{\psi_1(t)} = \frac{1}{4} + \frac{1}{\text{dim} E}\tr(I^E p_E) = \frac{1}{4} + \frac{3}{4}\frac{1}{4^E}.
\eea
where $1/4$ comes from the probability of selecting the first term of $\ket{\psi}$, which automatically guarantees EPR-states on the $E$-pairs, whereas, $3/4$ is the probability of selecting the second term, and $1/4^E$ is the chance that the said $E$-pairs are EPR for the term. Notice that the post-selection does not guarantee successful teleportation, as the second term of $\ket{\psi}$ selects a state orthogonal to the one we want on qubit $b$. However, the relative probability of selecting that term is low, even for $E$ which is not too large.

Given successful post-selection, the teleportation fidelity is given by
\bea
F_E = \frac{\bra{\psi(t)}p_E \ket{A}_b \bra{A}_b p_E \ket{\psi(t)}}{P_E} = \frac{1}{P_E}\left(\frac{1}{4} + \frac{3}{4}\bra{\psi_1(t)}p \ket{A}_b \bra{A}_b p \ket{\psi_1(t)} \right)
\eea
Without loss of generality, take $\ket{A}=\ket{0}$. Then we have
\bea
\bra{\psi_2(t)}p_E \ket{A}_b \bra{A}_b p_E \ket{\psi_2(t)} = \frac{1}{3} \bra{\epr}\sigma_1^z(t) p_E \sigma_1^z(t) \ket{\epr} = \frac{1}{3}\frac{1}{4^E}
\eea
The last step is obtained from the thermalization of the operator state $\sigma_1^z(t)\ket{\epr}$. 
Therefore, the fidelity given successful post-selection is 
\bea
F_E = 1 - \frac{2}{4^E+3},
\eea
which approaches unity exponentially fast as $E$ increases.

The calculation carries over to the teleportation protocol of the XY model presented in the main text, where the decoupled unitary dynamics $U\otimes U^*$ is replaced by a single forward evolution unitary $U = e^{-i H_{xy}t}$.
This is made possible by a special eigenstate of the XY model, which is related to the EPR state by single-qubit rotations. Instead of EPR pairs, the eigenstate $\ket{\eig}$ contain Bell pairs, $\ket{00}\pm\ket{11}$, that form a checkerboard pattern. 
The initial state of the teleportation is instead given by
\bea
\ket{\psi}= \frac{1}{2}\ket{\eig}\ket{A}_b +\sum\limits_{s=x,y,z} \frac{1}{2}\sigma_1^s\ket{\eig}\sigma_b^s\ket{A}_b
\eea
The first term is the eigenstate of the unitary and remains stationary. The second term is not, and thermalizes. Bob can perform Bell measurements and post-select the result to match the Bell state in $\ket{\eig}$ to select the first term. 
The post-selection probability and teleportation fidelity are the same as the Hayden-Preskill protocol.  

\section{Quantum Channel Details}
In this appendix we give further details on the quantum channel defined in Eq.~(6) of the main text and illustrate that $\ket{\text{EIG}}$ is a steady-state of the channel.
\subsection{Kraus operators of the quantum channel}
An initial density state, $\rho$ is subjected to a linear map, $\mathcal{M}$, which acts as,
\bea
\mathcal{M}[\rho] = \sum_{a}M_{a}\rho M_{a}^{\dagger}.
\eea
Where the Kraus operators $\{M_{a}\}$ are trace-preserving, ensured by the completeness relation, $\sum_{a}M^{\dagger}_{a}M_{a}=\mathbbm{1}$.
Recall that the quantum channel in the main text is written as,
\bea \label{eq:quantum_channel}
\mathcal{M}[\rho]
=K_{00}\rho K^{\dagger}_{00} + K_{11}\rho K^{\dagger}_{11}
+ \tr\left(K_{01}\rho K^{\dagger}_{01} \right)\rho_{0} + \tr\left(K_{10}\rho K^{\dagger}_{10}\right)\rho_{0}.
\eea
The operator, $K_s(T)=\bra{s}e^{-iH_{xy}T}\ket{I_{c\overline c}}$ with dimension $2^{N-2}$ where $s= \{00,11,01,10\}$ and the initial state, $\rho_0 = \ket{0\cdots 0}\bra{0\cdots 0}$. Each term represents each of four possible outcomes after measuring the center two qubits $q_c$ and $q_{\bar c}$.
\\

In the following, we explicitly construct the Kraus operators for this quantum channel and show that they satisfy the completeness relation.
The first two terms are already in the Kraus form.
The last two terms in Eq.~\eqref{eq:quantum_channel} can be expressed in Kraus form as follows, 
\bea
\tr\left(K_{01}\rho K^{\dagger}_{01} \right)\rho_{0} = \sum\limits_{n=0}^{\text{dim}-1} \ket{\psi_0} \bra{\psi_n} K_{01}\rho K_{01}^\dagger \ket{\psi_n}\bra{\psi_0} \\
\tr\left(K_{10}\rho K^{\dagger}_{10} \right)\rho_{0} = \sum\limits_{n=0}^{\text{dim}-1} \ket{\psi_0} \bra{\psi_n} K_{10}\rho K_{10}^\dagger \ket{\psi_n}\bra{\psi_0}
\eea
where $\ket{\psi_0}$ is the initial state $\ket{0\cdots 0}$ and $\ket{\psi_n}$ sums over the dimension of the computational basis $\text{dim} = 2^{N-2}$ from the remaining $2^{N-2}$ qubits, excluding $q_c$ and $q_{\bar c}$. Therefore the quantum channel in Eq.~\eqref{eq:quantum_channel} has rank $2^{N-1}+2$ and the $2^{N-1}+2$ Kraus operators are
\begin{equation}
    M_0 = K_{00}, \quad M_1 = K_{11}, \quad M_{2\rightarrow\text{dim}+1} = \ket{\psi_0}\bra{\psi_{0\rightarrow\text{dim}-1}}K_{01}, \quad M_{\text{dim}+2\rightarrow 2\text{dim}+2} = \ket{\psi_0}\bra{\psi_{0\rightarrow\text{dim}-1}}K_{10}
\end{equation}
Now we demonstrate the completeness relation of the Kraus operator.
Using $\sum_{n} \ket{\psi_n}\bra{\psi_n}=\mathbbm{1}$, one can show that
\bea
\sum_{a} M^\dagger_a M_a = K_{00}^\dagger K_{00} + K_{01}^\dagger K_{01} + K_{10}^\dagger K_{10} + K_{11}^\dagger K_{11} = \bra{I_{c\bar c}} e^{i H_{xy}T} e^{ - i H_{xy}T} \ket{I_{c\bar c}} = \mathbbm{1}.
\eea

\subsection{steady-state of the quantum channel}

Physically, the quantum channel $\mathcal{M}$ contains two steps:
\begin{enumerate}[1:]
\label{lst:protocol}
    \item (Evolution under $H_{\text{xy}}$): Evolve the global state for a time $T$ under $H_{\text{xy}}$.
    \item (Measurement and feedback): Measure the two-qubit state of $q_c$ and $q_{\bar c}$ in the computational basis. There are four possible outcomes, $\ket{00}$, $\ket{01}$, $\ket{10}$, $\ket{11}$. If the outcome is $\ket{11}$ or $\ket{00}$, rotate the two-qubit state to the EPR state. If the outcome is $\ket{01}$ or $\ket{10}$, reset the global state to $\ket{00\cdots00}$ then rotate the two-qubit state of $q_c$ and $q_{\bar c}$ to the EPR state.
\end{enumerate}
The important property of the quantum channel outlined in the main text is that it hosts $\ket{\text{EIG}}$ as a unique steady-state. 
This is because both steps leave $\ket{\eig}$ unchanged.
Importantly, the measurement in the second step disentangles the qubits $q_c$ and $q_{\bar c}$ from the rest of the system, which after the rotation is always in the EPR state after the second step in each iteration. Therefore, we can study the system's dynamics only by focusing on the density matrix of the remaining $N-2$ of the system. 
The quantum channel in turn encapsulates the two steps in the protocol into one neat mathematical expression. In the following, we break down how the steps act on the said eigenstate:
\begin{enumerate}[1:]
    \item (Evolution under $H_{xy}$) In this step, the entire state passes through unchanged, as it is an eigenstate of the XY model. 
    \item (Measurement of the qubits and resetting) In the eigenstate, $q_c$ and $q_{\bar c}$ form an EPR pair $\frac{1}{\sqrt 2}(\ket{00}+\ket{11})$. As a result, measuring the two qubits in the computational basis will result in $\ket{00}$ and $\ket{11}$, which is rotated back to the EPR state. Hence, this step also leaves $\ket{\eig}$ unchanged.
\end{enumerate}
Following this, the state $\ket{\eig}$ passes through $\mathcal{M}$ unchanged and therefore is a steady-state.
The uniqueness of the steady-state has been checked via exact diagonalization for the system sizes mentioned in the main text.
Once the uniqueness has been established, a finite gap guarantees that all the other components of the initial states will decay within a finite time, leaving behind the long-range entangled state. The larger the gap, the quicker it will converge.

\subsection{Other related quantum channels}
In this section, we present two closely related quantum channels to illustrate the design principle of the state-preparation protocol.

\begin{figure}
    \centering
    \includegraphics[width=0.31\columnwidth]{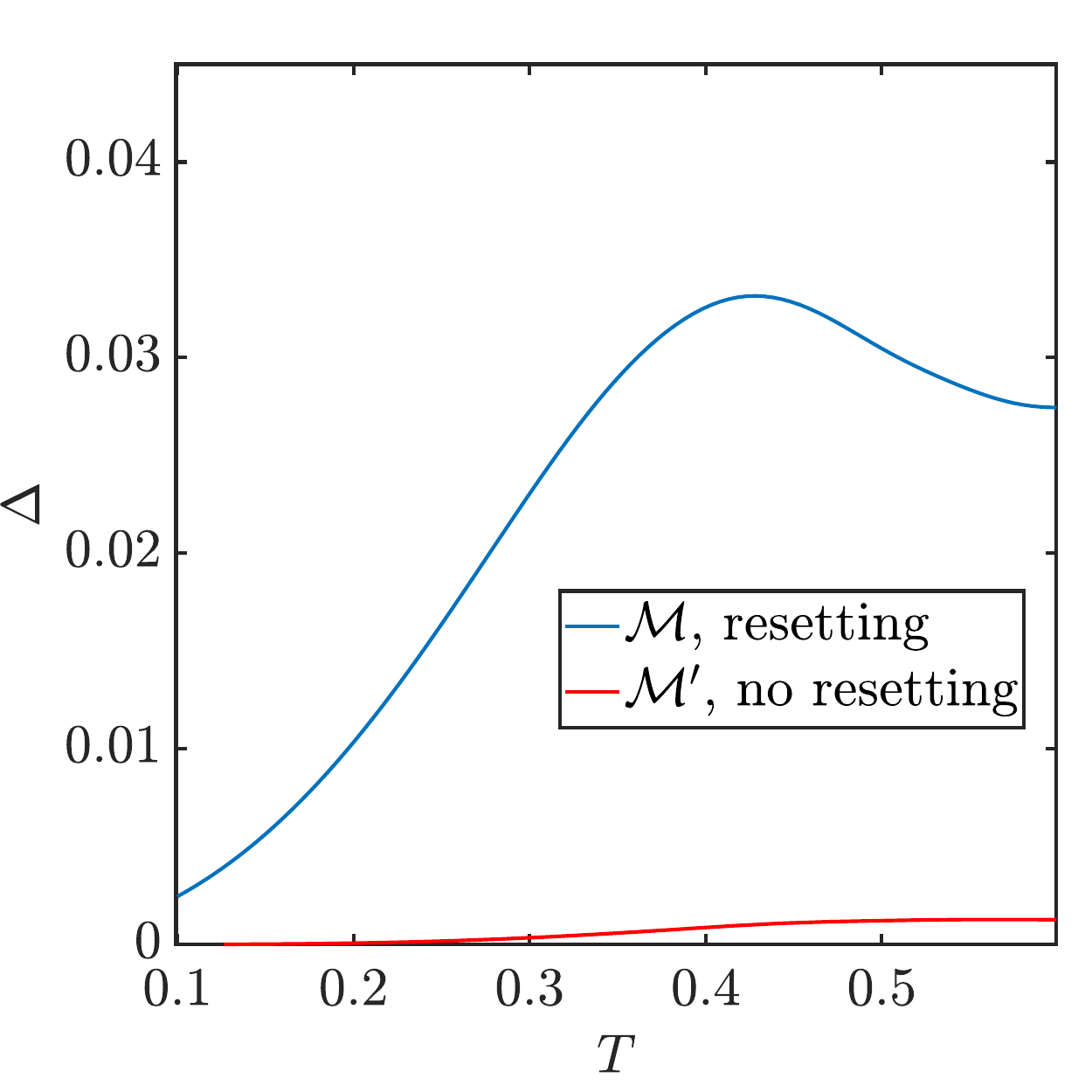}
    \includegraphics[width=0.31\columnwidth]{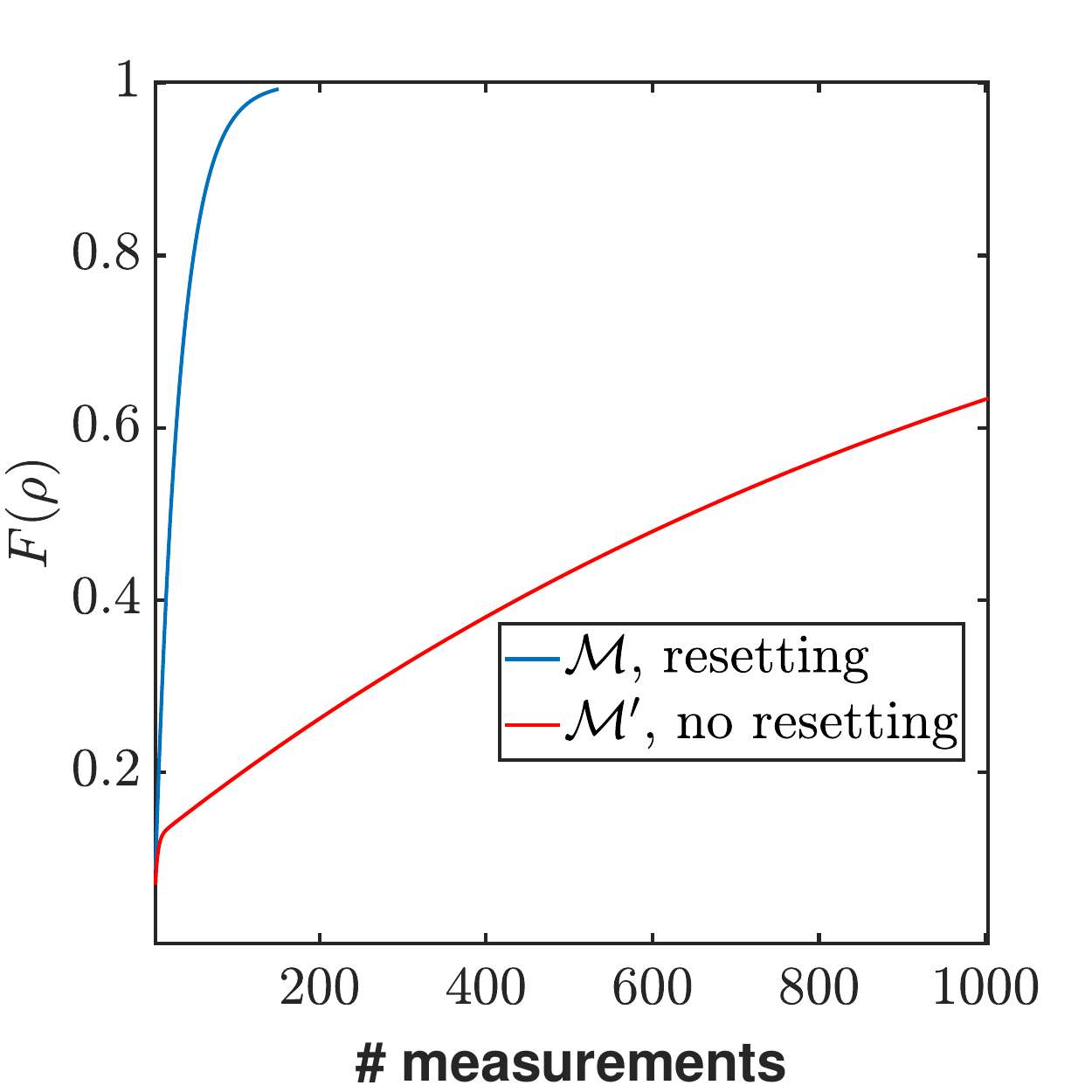}
    \includegraphics[width=0.31\columnwidth]{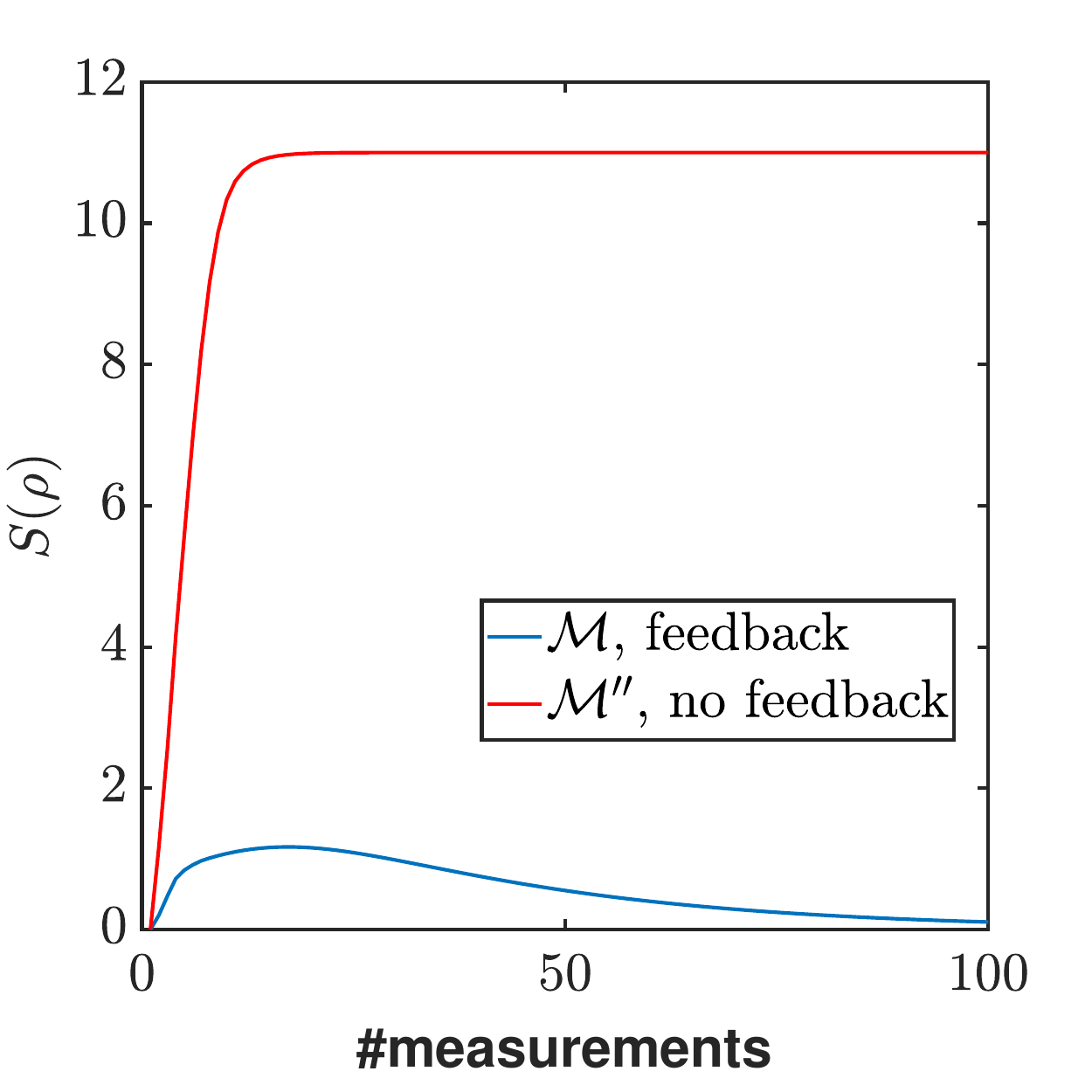}
    \caption{Comparison between different quantum channels for 12 qubits in the $6\times 2$ geometry. (a) The gaps of quantum channel $\mathcal{M}$ and $\mathcal{M'}$. The channel $\mathcal{M}$ includes the resetting procedure while $\mathcal{M}'$ does not. (b) The fidelity as a function of the number of iterations. An initial state converges to the targeted state much faster with the resetting procedure. We checked that the initial state under $\mathcal{M}'$ without resetting eventually reaches the targeted steady-state after $\sim 5000$ measurement.  (c) The entropy of the density matrix as a function of the number of iterations. The channel $\mathcal{M}''$ performs measurements in the Bell basis and does not include feedback control. As a result, the full mixed state is a steady-state of $\mathcal{M}''$ in addition to $\ket{\eig}$. Under $\mathcal{M}''$, a typical initial state thermalizes to the mixed steady-state instead of the targeted state. The entropy of the steady-state is 11 instead of the maximal entropy 12 because the protocol conserves the total parity. }
    \label{fig:otherprotocol}
\end{figure}

First, let us consider the quantum channel without resetting, denoted as $\mathcal{M}'$. The two steps are given as follows:
\begin{enumerate}[1:]
\label{lst:protocol}
    \item (Evolution under $H_{\text{xy}}$): Evolve the global state for a time $T$ under $H_{\text{xy}}$.
    \item (Measurement and feedback): Measure the two-qubit state of $q_c$ and $q_{\bar c}$ in the computational basis. Regardless of the outcome, rotate the two-qubit state to the EPR state. 
\end{enumerate}
One can verify that the state $\ket{\eig}$ is a steady of the quantum channel as both steps leave the state unchanged. The density matrix of the system is updated by
\begin{equation}
    \rho\rightarrow \mathcal{M}'(\rho) = K_{00}\rho K^{\dagger}_{00} + K_{01}\rho K^{\dagger}_{01} + K_{10}\rho K^{\dagger}_{10} + K_{11}\rho K^{\dagger}_{11}. 
\end{equation}
From exact diagonalization, $\ket{\eig}$ is still the unique steady-state of $\mathcal{M}'$, but the spectrum gap is far smaller than that of $\mathcal{M}$, shown in Fig.~\ref{fig:otherprotocol}(a), making the state preparation much less efficient. For the $6\times 2$ geometry, the system converges to the steady-state in 5000 iterations~(Fig.~\ref{fig:otherprotocol}(b)), compared with 150 using the protocol with resetting.  Furthermore, the system has to maintain coherence during all the iterations, in contrast with the protocol using $\mathcal{M}$, where the system only needs to maintain coherence between resettings.

Now we consider the second variant of the quantum channel, dubbed as $\mathcal{M}''$, where the measurements are in the Bell basis, and there is no feedback control. The steps are:
\begin{enumerate}[1:]
\label{lst:protocol}
    \item (Evolution under $H_{\text{xy}}$): Evolve the global state for a time $T$ under $H_{\text{xy}}$.
    \item (Measurement): Measure the two-qubit state of $q_c$ and $q_{\bar c}$ in the Bell basis, and there are four possible outcomes $\ket{I}$, $\ket{X}$, $\ket{Y}$  and $\ket{Z}$. 
\end{enumerate}
Again, one can check that $\ket{\eig}$ is a steady-state of the protocol since $q_c$ and $q_{\bar c}$ in $\ket{\eig}$ form an EPR and are not affected by the measurement. Here since the two-qubit state of $q_c$ and $q_{\bar c}$ is not in a definite state after each iteration, we need to describe the system by the density matrix of total $N$ qubits instead of $N-2$. 
The density matrix is updated via
\begin{equation}
\rho \rightarrow \mathcal{M''}(\rho) = \sum_{s=I, X, Y, Z}\tilde{K}_s \rho \tilde{K}_s^\dagger, \quad  \tilde K _s = \ket{S}_{c\bar c}\bra{S}_{c\bar c} e^{-i H_{xy}T}.
\end{equation}
For this protocol, although $\ket{\eig}$ is the steady-state, it is not unique. For example, one can check that the fully mixed state is also a steady-state. As a result, applying this protocol to an initial state is not guaranteed to converge to the targeted state $\ket{\eig}$. In fact, a typical initial state will converge to the fully mixed state and becomes featureless, confirmed by the dynamics of the thermal entropy $S(\rho)$ in Fig.~\ref{fig:otherprotocol}(c). 

In summary, the feedback control is required to set the special eigenstate $\ket{\eig}$ as the unique steady-state, and the resetting procedure is needed to make the protocol more efficient and reduce the required coherent time of the system.


\section{Numerical Implementation.}
In this work, we are restricted to small system sizes because both protocols require a large entanglement resource, eliminating the application of more sophisticated procedures such as, matrix product states.
We first outline the exact diagonalization procedure we utilized at the beginning stages of this work for small system sizes and then describe the pure state dynamics with Monte-carlo sampling.

For small system sizes, we may time-evolve the density matrix exactly under the quantum channel by first applying the operator-state duality,
\begin{equation}
    \rho = \sum_{ij}\rho_{ij}\ket{i}\bra{j} \rightarrow \ket{\rho}=\sum_{ij}\rho_{ij}\ket{i}\ket{j}.
\end{equation}
Before the mapping the density operator is a $2^{N}\times 2^{N}$ matrix that is then taken to a $4^{N}$ dimensional vector.
The state is then updated by the quantum channel as follows,
\begin{equation}
    \mathcal{M}\ket{\rho} = \sum_{a}\left(M_{a}\otimes M^{*}_{a} \right)\ket{\rho}.
\end{equation}
The quantum channel is now acting on an enlarged $4^{N}$ dimensional Hilbert space where $M_{a}^{*}$ denotes the complex conjugate with respect to the defined basis.
In matrix form, the quantum channel $\mathcal{M}$ has dimension $4^N \times 4^N$. Therefore, full diagonalization and obtaining the entire spectrum of the quantum channel is limited to system size $N=6$. 
In many cases, full diagonalization is not required.
For example, evolving the density matrix only requires applying the quantum channel to the density matrix.
The low-lying eigenvalues, whose real parts are near $1$, and the eigenstates can be obtained efficiently by the Lanczos algorithm which also only requires applying the quantum channel to the density matrix, which is done efficiently due to the quantum channel being sparse as a matrix.
The bottleneck in these approaches is to store the density matrix, a dense $2^N\times 2^N$ matrix.
As a result, the largest system size accessible with reasonable computation resources is $N=12$. 

The initial pure state used in our state-preparing protocol is $\ket{\psi_{0}}=\bigotimes_{i}\ket{0}_{i}$.
Applying a non-unitary operation, such as measurement, causes the density matrix to become mixed which then has to be stored.
This problem is circumvented by keeping the state pure and sampling over measurement outcomes to build a distribution to study the property of the mixed state, similar to what happens in real experiments.
Recall above how a quantum channel acts on an initial state $\ket{\psi}$, the resulting state is $\rho = \mathcal{M}[\ket{\psi}\bra{\psi}]$.
The resulting mixed state is written as the weighted average over all pure states,
\begin{equation}
   \rho =  \sum_a^n p_a \ket{\psi_a}\bra{\psi_a}
\end{equation}
The probability $p_a$ is $\langle \psi |M_a^\dagger M_a|\psi\rangle $ and $\ket{\psi_a}$ is the normalized state $M_a\ket{\psi}/p_a^{1/2}$.
Recall that $\rho$ is pure if $p_a = 1$ for some $a$.
Then any expectation value with respect to $\rho$ is obtained by sampling over the outcomes,
\begin{equation}
\text{tr}(\rho O)=\sum_a p_a \bra{\psi_a}O \ket{\psi_a}.
\end{equation}
Stochastic sampling is done using Monte Carlo: Draw a uniformly distributed random number $0<r<1$.
If 
\begin{equation}
    \sum\limits_{a=1}^m p_a < r < \sum\limits_{a=1}^{m+1} p_a,
\end{equation}
then $\ket{\psi_m}$ is chosen as the outcome state.

In our state-preparing protocol, the quantum channel is applied iteratively to the initial state. We have
\begin{equation}
    \rho=\mathcal{M}^n [\ket{\psi}\bra{\psi}] = \sum M_{a_n}\cdots M_{a_1}\ket{\psi}\bra{\psi}M^\dagger_{a_1} \cdots M^\dagger_{a_n}
\end{equation}
Then the density matrix can be written as
\begin{equation}
\rho = \sum p_{\{a\}} \ket{\psi_{\{a\}}}\bra{\psi_{\{a\}}}    
\end{equation}
The subscript $\{a\}$ stands for $a_n\cdots a_1$, which records each measurement outcome during the iterations and is called quantum trajectory.
The density matrix averages over all the quantum trajectories by Monte-Carlo sampling at each step.
Therefore, at a given iteration, only a pure state instead of a density matrix needs to be stored.
The quantum trajectory approach drastically increases the largest system size to $N=24$ with the drawback that one has to perform the simulation many times due to the stochastic nature of this approach. 
In practice, our largest system size is $N=18$, and every quantum trajectory converges to the targeted rainbow state, indicating that the steady-state of the quantum channel is the rainbow state.

\section{Special eigenstates of the XY model.}
In the main text, we focused on the non-equilibrium dynamics resulting from one special eigenstate of the XY model, given by
\bea
\ket{\eig}= \prod_{i+j\in \text{even}} \ket{I_{(i,j),(\overline{i},j)}} \prod_{i+j\in \text{odd}} \ket{Z_{(i,j),(\overline{i},j)}}.
\eea
In this section, we explicitly show that this state is an eigenstate of the XY model and also provide other closely related special eigenstates. To simplify the notation, in the following, we will use $X$, $Y$ and $Z$ to represent the three Pauli operators. The XY Hamiltonian on a 2D square lattice of size $L_x\times L_y$ is written as
\bea
H_{xy} = \sum_{\braket{rr'}} J_x X_r X_{r'} + J_y Y_r Y_{r'}
\eea
where $r$ and $r'$ are coordinates of nearest neighbor sites. We focus on even $L_x$ and denote the mirror reflection of coordinate $(i,j)$ as $(\overline i,j) \equiv (L_x+1-i,j)$. Let us first consider the term $X_{(i,j)}X_{(i+1,j)}$ on the vertical bonds. This term, acting on $\ket{\eig}$, affects two Bell pairs, one in $\ket{I}$ and the other one in $\ket{Z}$. We get
\bea
X_{(i,j)}X_{(i,j+1)} \ket{\eig} = - X_{(\overline{i},j)}X_{(\overline{i},j+1)} \ket{\eig}.
\eea
Therefore, the $XX$ terms on a vertical bond and its mirror partner cancel each other when acting on $\ket{\eig}$, the same for the $YY$ terms. This demonstrates that $\ket{\eig}$ is an eigenstate of all the terms on the vertical bonds with eigenvalue 0.

Now let us consider the $XX$ terms on the horizontal bond. We have
\bea
X_{(i,j)}X_{(i+1,j)} \ket{\eig} = - X_{(\overline{i},j)}X_{(\overline{i+1},j)} \ket{\eig}.
\eea
except for $i=L_x/2$, i.e., the central bonds on each row. Therefore, all terms except those on the central bonds cancel out when acting on $\ket{\eig}$. The terms on the central horizontal bonds obey the relations,
\bea
&X_{(L_x/2,j)}X_{(L_x/2+1,j)} \ket{\eig} = (-1)^{L_x/2+j} \ket{\eig},\\
&Y_{(L_x/2,j)}Y_{(L_x/2+1,j)} \ket{\eig} = -(-1)^{L_x/2+j} \ket{\eig}.\\
\eea
Upon which summing over $j$, we see that $\ket{\eig}$ is an eigenstate of the XY Hamiltonian with eigenvalue $-(J_x-J_y)(-1)^{L_x/2}\text{mod}(L_y,2)$.

There also exist other closely related eigenstates in the spectrum of the XY model. They are
\bea
\ket{\eig}_{IZ}= \prod_{i+j\in \text{even}} \ket{I_{(i,j),(\overline{i},j)}} \prod_{i+j\in \text{odd}} \ket{Z_{(i,j),(\overline{i},j)}},\\
\ket{\eig}_{ZI}= \prod_{i+j\in \text{even}} \ket{Z_{(i,j),(\overline{i},j)}} \prod_{i+j\in \text{odd}} \ket{I_{(i,j),(\overline{i},j)}},\\
\ket{\eig}_{XY}= \prod_{i+j\in \text{even}} \ket{X_{(i,j),(\overline{i},j)}} \prod_{i+j\in \text{odd}} \ket{Y_{(i,j),(\overline{i},j)}},\\
\ket{\eig}_{YX}= \prod_{i+j\in \text{even}} \ket{Y_{(i,j),(\overline{i},j)}} \prod_{i+j\in \text{odd}} \ket{X_{(i,j),(\overline{i},j)}}.
\eea
The first is the one discussed above and the other three are related to the first one by single-qubit rotations. Repeating the above analysis for the other three shows that they are indeed eigenstates with the corresponding eigenvalues; $(J_x-J_y)(-1)^{L_x/2}\text{mod}(L_y,2)$, $-(J_x+J_y)(-1)^{L_x/2}\text{mod}(L_y,2)$ and $(J_x+J_y)(-1)^{L_x/2}\text{mod}(L_y,2)$. They are degenerate for even $L_y$ but are distinct for odd $L_y$.

These eigenstates have \textit{maximal} entanglement entropy within each row and \textit{zero} entanglement entropy between rows. For even $L_y$, one can also write down four additional eigenstates containing the Bell states along the columns.

\bibliography{references}